# Toward a Science of Autonomy for Physical Systems: Service

Peter Allen
allen@cs.columbia.edu
Columbia University

Henrik I. Christensen
hic@cc.gatech.edu
Georgia Institute of Technology

## 1 Overview

A recent study by the Robotic Industries Association [2] has highlighted how service robots are increasingly broadening our horizons beyond the factory floor. From robotic vacuums, bomb retrievers, exoskeletons and drones, to robots used in surgery, space exploration, agriculture, home assistance and construction, service robots are building a formidable resume. In just the last few years we have seen service robots deliver room service meals, assist shoppers in finding items in a large home improvement store, checking in customers and storing their luggage at hotels, and pour drinks on cruise ships. Personal robots are here to educate, assist and entertain at home.  These domestic robots can perform daily chores, assist people with disabilities and serve as companions or pets for entertainment [1].  By all accounts, the growth potential for service robotics is quite large [2, 3].

Due to their multitude of forms and structures as well as application areas, service robots are not easy to define. The International Federation of Robots (IFR) has created some preliminary definitions of service robots [5]. A service robot is a robot that performs useful tasks for humans or equipment excluding industrial automation applications.  Some examples are domestic servant robots, automated wheelchairs, and personal mobility assist robots. A service robot for professional use is a service robot used for a commercial task, usually operated by a properly trained operator. Examples are cleaning robot for public places, delivery robot in offices or hospitals, fire-fighting robots, rehabilitation robots and surgery robots in hospitals. In this context an operator is a person designated to start, monitor and stop the intended operation of a robot or a robot system.

A degree of autonomy is required for service robots ranging from partial autonomy (including human robot interaction) to full autonomy (without active human robot intervention).  Therefore, in addition to fully autonomous systems service robots include systems, which may also be based on some degree of human robot interaction or even full tele-operation. In this context human robot interaction means information and action exchanges between human and robot to perform a task by means of a user interface. Service robots may consist of a mobile platform on which one or several arms are attached and controlled in the same mode as the arms of industrial robot.

---

[1] Contact: Ann Drobnis, Director, Computing Community Consortium (202-266-2936, adrobnis@cra.org).
For the most recent version of this essay, as well as related essays, please visit: cra.org/ccc/resources/ccc-led-white-papers

The market and economic impact of these robots is growing. A recent study [4] found that in 2013, about 4 million service robots for personal and domestic use were sold, 28% more than in 2012. The value of sales increased to US$1.7 billion. So far, service robots for personal and domestic use are mainly in the areas of domestic (household) robots, which include vacuum and floor cleaning, lawn-mowing robots, and entertainment and leisure robots, including toy robots, hobby systems, education and research. Handicap assistance robots have taken off to the anticipated degree in the past few years. In 2013 a total of about 700 robots were sold, up from 160 in 2012 - an increase of 345%. Numerous national research projects in many countries concentrate on this huge future market for service robots. In contrast to the household and entertainment robots, these robots are high-tech products. In 2013, it was estimated that 2.7 million domestic robots, including all types, were sold. The actual number might, however, be significantly higher, as the IFR survey is far from having full coverage in this domain. The value was about US$799 million, 15% higher than in 2012. Figure 1 captures some of this current and projected growth.

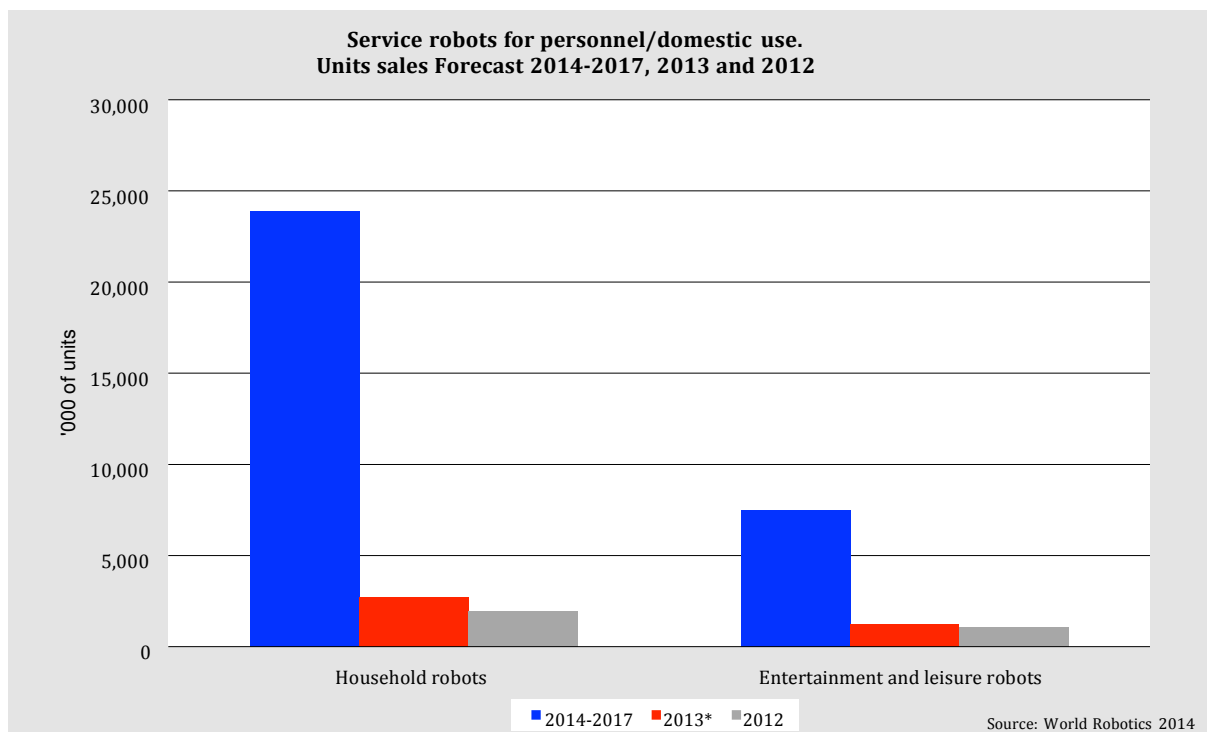

Figure 1: Forecast for service robots [4]

The SPARC partnership for robotics in Europe [6] is representative of the growing interest in robotics worldwide. SPARC's mission is based on the premise that robots are known to save costs, to improve quality and working conditions, and to minimize resources and waste. SPARC has estimated that from todays €22bn worldwide revenues, robotics industries are set to achieve annual sales of between €50bn and €62bn by 2020. [7].

## 2   Service Robots Impact and Application Domains

Below we highlight some application domains and research challenges for service robots. Many of these examples are drawn from the SPARC 2014 roadmap that is a comprehensive resource for autonomous service robots [7]. The emphasis here is on domestic service robots and less on professional applications.

### 2.1   Intelligent Robotic Appliances

Robotics technology has wide applicability within the domestic appliances market.  The addition of Robotics technology typically enhances products by extending functions through providing a degree of autonomy.  Over time there is a user expectation that robotic-based appliances will be able to complete many household tasks autonomously. Robotics technology has been applied to domestic appliances for over ten years, starting with pool cleaners, vacuum cleaners and lawn mowers. The market in these areas is now maturing and individual sales volumes are increasing. Dependability is critically important in the more advanced applications.  Appliances will need to be able to recognize failures and remain safe. End-users are always quick to recognize what works and what does not so fulfilling a genuine user need is essential. Moreover, systems are needed that are able to handle unexpected events in their environment, and recover seamlessly.

### 2.2   Assisted Living

Assisted living addresses the challenges of robotics technology support for independent living at all ages, social innovation and inclusion and ageing.  The main settings of this are the house, the town, and daily human-inhabited environments; on the other side the relative actors are mainly healthy persons.  The sub-domain of assisted living is closely related to the healthcare domain however its focus is on non-medical applications and on an ageing society. Non-medical consumer customers, such as individual citizens, elderly persons, their families and caregivers, define the market.  This domain addresses robotic solutions and technologies that aim to improve the quality of life by enriching the environments where humans live and work. These new technologies need to provide end-users with dependable, acceptable and sustainable support and assistance including, where necessary, individually tailored systems.

Technological challenges include human robot interaction, cognition and perception as well as mechatronics in order to create co-workers and companions able to provide appropriate levels of care and assistance. The primary abilities for this type of robot system are safe and intuitive interaction and configurability to each end users needs. In order to create such systems new design and development processes will be needed together with certification and testing able to provide guarantees of performance in everyday environments.  This requires also an integrative approach to science and engineering in order to overcome the bottleneck affecting traditionally engineered mechatronic modular systems that are in most cases built as simple sums of components.  The creation of such systems will require significant advances in system abilities particularly in dependability and safety and cognitive and interaction ability. Advances in these system abilities should be pursued together with the definition of new strategies and approaches aiming at endowing the new robots with highly integrated sensorimotor architectures and morphologies.

The core of providing assistive care is the development of sustainable systems designed around the human being that address the questions and challenges of the ageing society. Assistance in everyday tasks such as food preparation and cleaning are fundamental to extending the utility of the home for the elderly and infirm. This may ultimately result in a new ecosystem of sustainable consumer service-products. This will not be realized unless there is an increase in the acceptance of robots in society with respect to elderly care. Such a vision is still far in the future and within the medium term research horizon it is important to establish the underlying elements that will be required to deliver and deploy such systems and to develop trials and platforms able to benchmark and establish performance baselines.

These assistive care robots will eventually impact on a wide range of different functions. These can be characterized into a number of different areas:

• Domestic services, including cleaning, clearing, security and food preparation.

• Social companionship covering both social interaction, healthcare monitoring and telepresence.

• Extended living applications including personal hygiene, cognitive assistance and wellbeing, health monitoring and emergency assistance.

• Mobility both in terms of personal mobility assistance inside and outside of the home and transport over longer distances.

• Personal motivation to achieve as much as a person is capable of while providing protection and assistance

### 2.2.1 Personal Wellbeing Services

The demands of an ageing population and increased pressure on centralized healthcare mean that there is increased interest in services delivered at home. Robotics technology has the potential to act in a diagnostic and therapeutic role. Promoting wellbeing at home through improved exercise, diet and monitoring could have considerable health benefits and is preferable to the provision of central services.  There is the added benefit that such systems are able to carry out multiple functions and provide continuous monitoring in a home setting, as opposed to sporadic checks in hospital outpatient departments. In the future it is possible the robots may be able to assist in cognitive and mental wellbeing by providing cognitive support even in assessing and reducing stress. Key to the success of these devices is the development of acceptable and effective sensing systems. Many physiological measurements require physical contact and measuring emotional state or behavioral traits, critical for the diagnosis of progressive conditions, requires continual monitoring and interpretation.  For acceptance of robots to occur, there are three basic requirements:  motivation for using the robot, sufficient ease of use, and comfort with the robot physically, cognitively and emotionally.

### 2.2.2 Robots for Personal Mobility

Mobility is a key element in the maintenance of a healthy life and a lack of mobility contributes to the onset of many age related health issues. Robotics technology has the potential to provide a wide range of different types of mobility aids from assistance in standing and sitting, to preventing falls, and helping with personal hygiene. Autonomous transport and assistance in mobility outside of the home is critical to extending social integration and maintaining a healthy life. This can include the entire spectrum from intelligent wheelchairs to self-driving cars, including semi-autonomous driving aids to assist people with hearing, visual, motor or other disabilities.  The development of mobility aids for walking that increase confidence in moving over longer distances is also an important objective. Smart mobility aids may also be enhanced through wider connection to sources of data in the cloud to ensure safety and the delivery of localized services.

Of critical importance to the utilization of such devices is their ergonomic acceptability coupled to the cost of deployment and ethical and legal issues, especially legal liability. Systems that are justified though cost saving will need to demonstrate continued and sustained performance over extended periods of time. Validating and certifying systems will also be critical to acceptability. Which this type of system there area also ethical and societal consequences to their use and deployment, particularly if this is wide spread. Public engagement and debate will be an essential apart of developing such systems.

### 2.2.3 Robots for the Hospitality Sector

Robotic assistants are poised to offer a wide range of functions and services in the hospitality sector.  Already, companies are making inroads with room service delivery robots [8] and robot concierge services [9]. This is a potentially fast growing sector of industry that can be exploited by robotic  technology.  Even though these robots need to be able to operate autonomously, they still need to interface with humans during their work. This puts an extra emphasis on HRI social interaction, and the success of these robots will be highly dependent on how well accepted they are by their human customers. Clearly, this dimension of HRI needs to be a focus of future research to allow robots to seamlessly be integrated into tasks spaces currently populated by humans.

Safety as well is a driving research focus for this class of robots as they operate autonomously in complex and dynamic environments.

## 3  Key Technologies and Areas of Research

### 3.1  Hardware Challenges

Advances in many core robotics technology areas are needed to realize this next generation of low- cost, robust robots, many of which will need both mobility and dexterous manipulation capabilities. Bringing the cost of the hardware down and making it easy to manufacture and repair will be a driver in this sector. Some example areas of research include:

• Strong lightweight materials

• Extended battery life and power management

• Mobile bases (including stair climbing)

• Compliant actuation

• Multi-fingered hands and high-payload arms

• New sensor technology including smart skin sensing for mobile manipulators, improved 2D and 3D vision sensors, force/torque sensing for safety and compliance, and better audition for HRI social interaction.

### 3.2  Software Challenges

There are many challenges in software that need to be addressed to make these robots safe, functional, trustworthy and reliable. Many of these technologies are already part of core Robotics research.  However, we note that making these systems fully autonomous will require new re- search thrusts and new capabilities to surpass human-in-the-loop planning, reasoning, control and learning. A partial list includes:

• Mapping: perceiving and understanding complex, dynamic 3D environments. Integrating semantic reasoning into mapping of objects and spaces.

• Planning: Smooth motion planning, collision avoidance, error recovery, global path planning.

• Human-Robot Interfaces: Improved Natural Language Processing, integration of social cues in HRI, emotional state understanding, interaction with multiple agents, user-friendly design.

- Cognition: recognition of user state, adapting templates/learning for novel situations, semantic reasoning.

- Learning: Understanding and learning human preferences and adaptation over time. Integrating new learned skills into the task space.

- Control: Integrating multi-sensor feedback into real-time control loops.

- Safety and Reliability: fail-safe execution, redundancy, HRI interaction to understand novel and critical situations.

## 4 Summary

In summary, the service robot sector has a large potential positive impact on our society. It can create a multi-billion dollar marketplace, free people from tasks that are time-consuming and un- interesting, and assist an ageing and disabled population. The current state of the art in this area consists of many prototype and special purpose robots that are far from robust, safe, capable and trustworthy. However, the promise of these robots is very real and also realizable in a 5-10 year horizon with proper focus and funding.

## References

[1]http://www.entrepreneur.com/article/245301
[2]http://www.robotics.org/content-detail.cfm?content_id=4925
[3]http://www.worldrobotics.org/uploads/media/Executive_Summary_WR_2014_02.pdf
[4] http://www.ifr.org/uploads/media/Executive_Summary_WR_2014.pdf
[5] http://www.ifr.org/service-robots/
[6] http://sparc-robotics.eu/about/
[7] http://sparc-robotics.eu/wp-content/uploads/2015/02/Multi-Annual-Roadmap2020-ICT-24-Rev-B-full.pdf
[8] http://www.savioke.com
[9] http://www.technewstoday.com/21435-robot-concierge-and-staff-japanese-hotel-making-it-a- reality/

*For citation use*: Allen P. & Christensen H. (2015). *Toward a Science of Autonomy for Physical Systems: Service*: A white paper prepared for the Computing Community Consortium committee of the Computing Research Association. http://cra.org/ccc/resources/ccc-led-whitepapers/

*This material is based upon work supported by the National Science Foundation under Grant No. (1136993). Any opinions, findings, and conclusions or recommendations expressed in this material are those of the author(s) and do not necessarily reflect the views of the National Science Foundation.*